\documentclass[twocolumn,showpacs,preprintnumbers,amssymb,prl]{revtex4-1}
\usepackage{graphicx}
\usepackage{dcolumn}
\usepackage{color}
\usepackage{bm}

\def \be{\begin{equation}}
\def \ee{\end{equation}}
\def \ben{\begin{eqnarray}}
\def \een{\end{eqnarray}}

\begin{document}

\title{Masking singularities with $k-$essence fields in an emergent gravity metric}
\author{Debashis Gangopadhyay}
\altaffiliation{debashis@bose.res.in}
\affiliation{S. N. Bose National Centre for Basic Sciences,
Salt Lake, Kolkata 700 098, India}
\author{Goutam Manna}
\altaffiliation{goutammanna.pkc@gmail.com}
\affiliation{Department of Physics, Prabhat Kumar College, Contai,Purba Medinipur-721401,India}
\author{Sourav Sen Choudhury }
\altaffiliation{senchoudhurys@gmail.com}
\affiliation{Department of Theoretical Physics, Ramkrishna Mission Vivekananda University,PO Belur Math,Howrah-711202,India}

\begin{abstract}
It is known that dynamical solutions of the $k$-essence equation of
motion change the metric for the perturbations around these
solutions and the perturbations propagate in an  emergent spacetime
with metric $\tilde G^{\mu\nu}$ different from the gravitational 
metric $g^{\mu\nu}$. We show that for observers travelling with 
the perturbations, there exist homogeneous field configurations for 
the lagrangian $L=[{1\over 2}g^{\mu\nu}\nabla_{\mu}\phi\nabla_{\nu}\phi]^{2}$
for which a singularity in the gravitational metric $g^{\mu\nu}$ can be 
masked or hidden for such observers. This is shown for the Schwarzschild 
and the Reissner-Nordstrom metrics.
\end{abstract}

\maketitle

{\bf 1.Introduction}

Present day observations have established that the universe consists of 
roughly 25 percent dark matter, 70 percent dark energy , about 4 percent 
free hydrogen and helium with the remaining one percent consisting of stars, 
dust, neutrinos and heavy elements. Actions with non-canonical kinetic terms 
have been shown to be strong candidates for dark matter and dark energy. 
A theory with a non-canonical kinetic term was first proposed by Born and 
Infeld in order to get rid of the infinite self-energy of the electron 
\cite{born}. Similar theories were also studied in \cite{heisenberg,dirac}.
Cosmology witnessed these models first in the context of scalar fields 
having non-canonical kinetic terms which drive inflation. Subsequently  
$k-$essence models of dark matter and dark energy were also 
constructed \cite{armen1,armen2,armen3,chiba,armen4,ark1,ark2,cal}.
Effective field theories arising from string theories also have 
non-canonical kinetic terms \cite{callan,gib1,gib2,sen}. 

An  approach to understand the origins of dark matter and dark energy involve
setting up lagrangians for what are known as $k-$essence fields
in a Friedman-Robertson-Walker metric with zero curvature constant.
In one approach
\cite{scherrer} it is possible to unify the dark matter and dark energy
components into a single scalar field model with the scalar field $\phi$ having a
non-canonical kinetic term. These scalar fields are the $k-$essence fields
mentioned above.
The general form of the lagrangian for these $k-$essence models is assumed to
be a function $F(X)$ with $X={1\over 2}g^{\mu\nu}\nabla_{\mu}\phi\nabla_{\nu}\phi$, 
and do not depend explicitly
on $\phi$ to start with. In \cite{scherrer}, $X$ was shown to satisfy a general
scaling relation,{\it viz.} $X ({dF\over dX})^{2}=Ca(t)^{-6}$ with $C$ a
constant (similar expression was also derived in \cite{chimento}).

Recently a lagrangian for the  $k-$essence field has been set up \cite{gango1} 
in a homogeneous and isotropic universe where
there are two generalised coordinates $q(t)= ln~a(t)$ ($a(t)$ is the scale factor)
and a scalar field $\phi(t)$ with a complicated polynomial interaction between them.
In the lagrangian, $q$ has a standard kinetic term while $\phi$ does not have a
kinetic part and occurs purely through the interaction term. \cite{gango1}
incorporates the scaling relation of \cite{scherrer}.
In \cite{gango2} questions regarding the amplitude of a scale factor 
at some epoch evolving to a different value at a later epoch was 
addressed for the above lagrangian at times close to the big bang 
(very small scale factor). As the scale factor is inversely proportional 
to the temperature at a particular epoch, these amplitudes  provided an 
estimate of quantum fluctuations of the temperature.

Relativistic field theories with canonical kinetic terms differ from 
lagrangian theories of $k$-essence in that non-trivial dynamical solutions 
of the $k$-essence equation of motion not only spontaneously break 
Lorentz invariance but also change the metric for the perturbations 
around these solutions \cite{babi1}. The perturbations propagate 
\cite{babi1,vis} in an emergent spacetime with metric 
$\tilde G_{\mu\nu}$ different from (and also not 
conformally equivalent to) the gravitational metric $g_{\mu\nu}$.

Now $g_{\mu\nu}$ can contain physical singularities. The motivation 
of this work is to investigate whether scenarios can be constructed  
where the singularity in $g_{\mu\nu}$ can be "masked" to observers 
travelling piggy-back on the perturbations of the $k$-essence 
scalar fields. Lagrangians for k-essence scalar fields have the 
general form \cite{babi1},  
$L(X,\phi)= V(\phi) + K_{1}(\phi) X + K_{2}(\phi) X^{2} + ...$ In \cite{babi1}
various scenarios have been described for this lagrangian including those 
linear in $X$. We show here that a simple model lagrangian 
quadratic in $X$,{\it viz.} 
$ L = X^{2}$ has $k$-essence field  configurations (for 
both the Schwarzschild and the Reissner-Nordstrom metrics) for which the singularities can be  
masked for observers sitting on the scalar field perturbations.
The plan of the paper is as follows. In Section 2 a brief summary is given of  emergent gravity 
concepts as developed in \cite{babi1}. In Section 3 the Schwarzschild metric is considered while  
Section 4 deals with the Reissner-Nordstrom case. Section 5 is the conclusion.

\section{Emergent Gravity}

Consider the $k-$essence scalar field $\phi$ minimallly coupled to the gravitational
field $g_{\mu\nu}$. Then the $k-$essence action  is
+$$
S_{k}[\phi,g_{\mu\nu}]= \int d^{4}x {\sqrt -g} L(X,\phi)
\eqno(1)$$
where $X={1\over 2}g^{\mu\nu}\nabla_{\mu}\phi\nabla_{\nu}\phi$ and $\nabla_{\mu}$ means the
covariant derivative asociated with the metric $g_{\mu\nu}$. The total action describing the
dynamics of $k-$essence and gravity is
$$
S[\phi,g_{\mu\nu}]= \int d^{4}x {\sqrt -g}[ -{1\over 2} M^{2}_{\mathrm Pl}R + L(X,\phi)]
\eqno(2)$$
where $R$ is the Ricci scalar and $M_{\mathrm Pl}$ the reduced Planck mass.
The energy momentum tensor for the $k-$essence field is (with $L_{\mathrm X}= {dL\over dX},~~ L_{\mathrm XX}= {d^{2}L\over dX^{2}}, 
L_{\mathrm\phi}={d\phi\over dX}$)
$$
T_{\mu\nu}= {2\over \sqrt {-g}}{\delta S_{k}\over \delta g^{\mu\nu}}= L_{X}\nabla_{\mu}\phi\nabla_{\nu}\phi - g_{\mu\nu}L
\eqno(3)$$
and the equation of motion for the $k-$essence field is
$$-{1\over \sqrt {-g}}{\delta S_{k}\over \delta \phi}= \tilde G^{\mu\nu}\nabla_{\mu}\nabla_{\nu}\phi +2XL_{X\phi}-L_{\phi}=0
\eqno(4a)$$
where the effective metric $\tilde G^{\mu\nu}$ is
$$\tilde G^{\mu\nu}= L_{X} g^{\mu\nu} + L_{XX} \nabla^{\mu}\phi\nabla^{\nu}\phi
\eqno(5a)$$
and is physically meaningful only when
$$1+ {2X  L_{XX}\over L_{X}} > 0$$
i.e the sound speed $c_{\mathrm s}=(1+ {2X  L_{XX}\over L_{X}})^{-1/2}$ is
a real quantity.
When this condition holds everywhere the effective metric $\tilde G^{\mu\nu}$ determines the
characteristics for $k-$essence \cite{armen4,gib3,gib4,ren} .For the non-trivial configurations
of the $k-$ essence field $\partial_{\mu}\phi\neq 0$ and  $\tilde G^{\mu\nu}$ is not conformally
equivalent to $g^{\mu\nu}$. So the characteristics are different from canonical
scalar fields whose lagrangians are linear in $X$. The characteristics determine the
local causal structure of the spacetime at every point of the manifold. So the local causal
structure for the $k-$essence field is different from those ones defined
by $g^{\mu\nu}$.

Making a conformal transfornmation
$G^{\mu\nu}\equiv {c_{\mathrm s}\over L_{X}^{2}}\tilde G^{\mu\nu}$ and
using the expression for $T_{\mu\nu}$ from equation $(3)$ one can
write the inverse of the metric $G^{\mu\nu}$ as
$$G_{\mu\nu}= {L_{X}\over c_{\mathrm s}}g_{\mu\nu}
-c_{\mathrm s}L_{XX}\nabla_{\mu}\phi\nabla_{\nu}\phi\eqno(5b)$$
We will be using this expression for the effective metric in all
that follows. Also note that after this conformal transformation,
if we further assume that $L$ is not an explicit function of $\phi$
then the equation of motion $(4a)$ is replaced by ;
$$-{1\over \sqrt {-g}}{\delta S_{k}\over \delta \phi}
= {L_{X}^{2}\over c_{\mathrm s}}G^{\mu\nu}\nabla_{\mu}\nabla_{\nu}\phi=0
\eqno(4b)$$

\section{The Schwarzschild solution}

The Schwarzschild metric is given by  ($r_{s}= 2GM/c^{2}\equiv 2GM $, taking $c=1$)
$$
ds^{2}=(1- {r_{\mathrm s}\over r}) dt^{2} 
- (1- {r_{\mathrm s}\over r})^{-1} dr^{2}\nonumber\\
- r^{2} (d\theta^{2} + sin^{2}\theta d\Phi^{2})
\eqno(6)$$
and the emergent metric components $ G_{\mu\nu}$ are related to the
Schwarschild metric components
$g_{\mu\nu}$ by $(5b)$. Therefore for $L= X^{2}$,and
assuming the $k-$essence field to be spherically symmetric
i.e.$\phi\equiv\phi(r,t)$ one has
$$G_{00}=\bigl(1- {r_{\mathrm s}\over r}\bigr) 2{\sqrt 3}X 
-{2\over {\sqrt 3}}\bigl({\partial\phi\over\partial t}\bigr)^{2}$$
$$G_{11}=-\bigl(1- {r_{\mathrm s}\over r}\bigr)^{-1} 2{\sqrt 3}X 
- {2\over{\sqrt 3}}\bigl({\partial\phi\over\partial r}\bigr)^{2}$$
$$G_{22}=2{\sqrt 3}Xg_{22}=-2{\sqrt 3}Xr^{2}$$
$$G_{33}=2{\sqrt 3}Xg_{33}=-2{\sqrt 3}Xr^{2}sin^{2}\theta\eqno(7)$$
$$G_{01}= G_{10}=-{2\over {\sqrt 3}}{\partial\phi\over\partial t}{\partial\phi\over\partial r}\eqno(8)$$
All the other $G_{\mu\nu}$ are zero.

Assume $\phi(r,t)$ to be of the form
$\phi_{\mathrm s}(r,t) =\phi_{\mathrm 1s}(r) + \phi_{\mathrm 2s}(t)$.
Then
$$G_{00}=\bigl(1- {r_{\mathrm s}\over r}\bigr) 2{\sqrt 3}X 
- {2\over{\sqrt 3}} \bigl({d\phi_{\mathrm 2s}\over dt}\bigr)^{2}$$ 
$$G_{11}=-\bigl(1- {r_{\mathrm s}\over r}\bigr)^{-1} 2{\sqrt 3}X 
- {2\over {\sqrt 3}}\bigl({d\phi_{\mathrm 1s}\over dr}\bigr)^{2}
\eqno(9)$$
$$G_{01}=G_{10}
=-{2\over{\sqrt 3}}\dot\phi_{\mathrm 2s}\phi_{\mathrm 1s}'
\eqno(10)$$
where "dot" denotes differentiation with respect to time and the "prime" is
differentiation with respect to $r$.
As we are concerned only with the singularity structure of the metrics we are not discussing
the $G_{22}$ and $G_{33}$ components as $g_{22}$ and $g_{33}$ are well behaved
for $r\rightarrow 0$.

We assume that
$L_{\mathrm X}~~;
~~L_{\mathrm XX}~~;
~~L_{XX}\bigl({\partial\phi\over\partial t}\bigr)^{2}~~;
~~L_{XX}\bigl({\partial\phi\over\partial r}\bigr)^{2}$
are all well behaved quantities for $r\rightarrow 0$.
All these conditions hold true in the above equations if we
also assume that $({\partial\phi_{\mathrm 1s}\over\partial r})$ is well behaved for
$r\rightarrow 0$.
We shall consider only physical singularities.
Here these occur at $r=0$.
The singularities at $r=r_{\mathrm s}$ are coordinate
singularities and these can always be removed by some
coordinate transformations and we are not considering them.

Note that at $r=0$, the second terms on the
{\it r.h.s.} of $(9)$ and $(10)$ are well behaved
as per our assumptions. Therefore good behaviour
of $G_{00}$ and $G_{11}$ at $r=0$
is guaranteed if there exist two functions
$f_{1}(r), f_{2}(r)$ such that both these functions
are well behaved at $r=0$ and
$$ (1- {r_{\mathrm s}\over r}) 2{\sqrt 3}X = f_{\mathrm 1}(r)~~;
~~(1- {r_{\mathrm s}\over r})^{-1} 2{\sqrt 3}X = f_{\mathrm 2}(r)
\eqno(11)$$
These equations imply that
$f_{1}(r)=f_{2}(r)(1- {r_{\mathrm s}\over r})^{2}$.
It is readily seen that for $X$ to
be well behaved as $r\rightarrow0$ and for the two equations
in $(11)$  to be consistent one possibility is
$f_{1}(r)= constant=1$ and
$f_{2}(r)= (1- {r_{\mathrm s}\over r})^{-2}$.
Then $X$ is well behaved at $r=0$. So
$$G_{00}=1 - {2\over{\sqrt 3}}(\dot\phi_{\mathrm 2s})^{2}~~;~~
G_{11}= -(1- {r_{\mathrm s}\over r})^{-2} -{2\over{\sqrt 3}} (\phi_{\mathrm 1s}')^{2}
\eqno(12)$$
At $r=0$ both $G_{00}~~,~~ G_{11}$ are well behaved and
$$X= {1\over 2{\sqrt 3}}{1\over (1- {r_{\mathrm s}\over r})}
\eqno(13)$$
With our assumption regarding the form of $\phi(r,t)$ , this leads to
$$(\dot\phi_{\mathrm 2s}(t))^{2} 
= {1\over{\sqrt 3}} + (1- {r_{\mathrm s}\over r})^{2})(\phi_{\mathrm 1s}'(r))^{2} = k
\eqno(14) $$
where $k$ is a constant. Note that $(12)$ and $(14)$ imply that if
the sign of (temporal component) $G_{00}$ has to remain positive
w.r.t. (spatial components) $G_{11},G_{22},G_{33}$, then
$G_{00} > 0$. This means $k < {3\over 4}=0.75$. Only these
values of $k$ are allowed.

We now discuss possible solutions to this equation. Note that

{\bf  Case 1, $k=0$}

We rule out taking $k=0$ because then
$\dot\phi_{\mathrm 2s}(t)= 0$ which means that the $k-$essence scalar
field does not have any kinetic energy. This  violates the basic premise of $k-$essence
where the kinetic energy drives the accelerated expansion.

{\bf  Case 2, $k={1\over{\sqrt 3}}=0.5773<0.75$}

Now we have  ,
$\phi_{\mathrm 1s}'= 0$ and $\dot\phi_{\mathrm 2s}=(3)^{-1/4}$ so that
$$\phi_{\mathrm 1s}(r)= c_{1}~~;~~\phi_{\mathrm 2s}(t) = (3)^{-1/4}t + c_{2}
\eqno(15)$$
where $c_{1}, c_{2}$ are constants. Now
$X= {1\over 2{\sqrt 3}}{1\over (1- {r_{\mathrm s}\over r})}$ and
$G_{00}= 1- {2\over{\sqrt 3}}{1\over{\sqrt 3}} = {1\over 3}\neq g_{00}~;~
G_{11}= -{1\over (1- {r_{\mathrm s}\over r})^{2}}\neq g_{11}~;~
G_{22}=2{\sqrt 3}Xg_{22}~;~ G_{33}=2{\sqrt 3}Xg_{33}~;~
G_{01}= G_{10}=0$. All the other off-diagonal components are also
zero. So the emergent metric without any singularity at $r=0$ is
$$ G_{\mu\nu} = \left(\begin{array}{cccc}
{1\over 3} & 0 & 0 & 0\\
0 & {-1\over (1- {r_{\mathrm s}\over r})^{2}} 
 & 0 & 0\\
0 & 0 & {-r^{2}\over (1- {r_{\mathrm s}\over r})} & 0\\
0 & 0 & 0 & {-r^{2} sin^{2}\theta\over (1- {r_{\mathrm s}\over r})} \\
\end{array}\right) \eqno(16)$$
It is straightforward to see from eqs.$(16)$ that
$G_{\mu\nu}$ and $g_{\mu\nu}$ are not conformally equivalent.

Therefore there exist homogeneous (i.e. independent of $r$)
$k-$essence scalar field configurations, {\it viz.}  $\phi (r,t)=  c_{1} + (3)^{-1/4}t + c_{2}$,
that can give rise to an emergent gravity metric where the singularity in the
gravitational metric $g_{\mu\nu}$ is masked for observers riding on the scalar
field perturbations.These configurations also satisfy the emergent gravity
equations of motion $(4b)$ as is easily seen:
${L_{X}^{2}\over c_{\mathrm s}}[G^{00}\partial_{0}^{2}\phi_{\mathrm 2s} 
+ G^{11}(\partial_{1}^{2}\phi_{\mathrm 1s} 
-\Gamma_{11}^{1}\partial_{1}\phi_{\mathrm 1s})
+G^{01}\nabla_{0}\nabla_{1}\phi 
+G^{10}\nabla_{1}\nabla_{0}\phi]= 0$. The first two terms within third brackets
vanish because $\phi_{\mathrm 2s}$ is linear in $t$ and $\phi_{\mathrm 1s}$ is
a constant. The last two terms vanish because $G^{01}\nabla_{0}\nabla_{1}\phi 
+G^{10}\nabla_{1}\nabla_{0}\phi=G_{01}\nabla^{0}\nabla^{1}\phi 
+G_{10}\nabla^{1}\nabla^{0}\phi$ and $G_{01}= G_{10}=0$.

\section{The Reissner-Nordstrom black hole}

For a static charged black hole  with charge $Q$ the metric is the Reissner-Nordstrom metric:
$$ds^{2}=(1- {r_{\mathrm s}\over r} + {r_{\mathrm Q}^{2}\over r^{2}}) dt^{2}
- (1- {r_{\mathrm s}\over r} + {r_{\mathrm Q}^{2}\over r^{2}})^{-1} dr^{2}\nonumber\\
- r^{2} (d\theta^{2} + sin^{2}\theta d\Phi^{2})
\eqno(17)$$
with $r_{\mathrm Q}^{2}= GQ^{2}/4\pi\epsilon_{0}c^{4}\equiv GQ^{2}/4\pi\epsilon_{0}$
taking $c=1$.
We now carry out an exactly similar analysis
as before for the same lagrangian $L= X^{2}$ and assume the solutions for
$\phi$ to be of the form
$\phi_{\mathrm n}(r,t) =\phi_{\mathrm 1n}(r) + \phi_{\mathrm 2n}(t)$.
Then
$$G_{00}=\biggl(1- {r_{\mathrm s}\over r}+ {r_{\mathrm Q}^{2}\over r^{2}}\biggr)2{\sqrt 3}X 
- {2\over {\sqrt 3}}\biggl({d\phi_{\mathrm 2n}\over\partial t}\biggr)^{2}$$
$$G_{11}=-\biggl(1- {r_{\mathrm s}\over r}+{r_{\mathrm Q}^{2}\over r^{2}}\biggr)^{-1} 2{\sqrt 3}X 
-{2\over{\sqrt 3}}\biggl({d\phi_{\mathrm 1n}\over\partial r}\biggr)^{2}
\eqno(18a)$$
$$G_{01}=G_{10}
=-{2\over {\sqrt 3}}\dot\phi_{\mathrm 2n}\phi_{\mathrm 1n}';~
G_{22}=2{\sqrt 3}X g_{22}~;~G_{33}=2{\sqrt 3}X g_{33}
\eqno(18b)$$
All the other $G_{\mu\nu}$ are zero.

As before at  $r=0$, the second terms on the
{\it r.h.s.} of $(18a)$ are well behaved
as per our assumptions. So good behaviour
of $G_{00}$ and $G_{11}$ at $r=0$
is guaranteed if there exist two functions
$g_{1}(r), g_{2}(r)$ such that both these functions
are well behaved at $r=0$ and
$$ (1- {r_{\mathrm s}\over r}+{r_{\mathrm Q}^{2}\over r^{2}})2{\sqrt 3}X = g_{\mathrm 1}(r)~~;
~~(1- {r_{\mathrm s}\over r}+{r_{\mathrm Q}^{2}\over r^{2}})^{-1}2{\sqrt 3}X = g_{\mathrm 2}(r)
\eqno(19)$$
These equations imply that
$g_{1}(r)=g_{2}(r)(1- {r_{\mathrm s}\over r}+{r_{\mathrm Q}^{2}\over r^{2}})^{2}$.
For $X$ to be well behaved as $r\rightarrow0$ and for consistency
one possibility is
$g_{1}(r)= constant=1$ and
$g_{2}(r)= \biggl(1- {r_{\mathrm s}\over r}+ {r_{\mathrm Q}^{2}\over r^{2}}\biggr)^{-2}$.
Then $X$ is well behaved at $r=0$. So
$$G_{00}=1 -{2\over{\sqrt 3}}(\dot\phi_{\mathrm 2n})^{2}~~;~~
G_{11}= -(1- {r_{\mathrm s}\over r}+{r_{\mathrm Q}^{2}\over r^{2}})^{-2} -{2\over{\sqrt 3}}
(\phi_{\mathrm 1n}')^{2}
\eqno(20)$$
At $r=0$ both $\tilde G_{00}~~,~~\tilde G_{11}$ are well behaved and
$$X= {1\over 2{\sqrt 3}}{1\over (1- {r_{\mathrm s}\over r}+{r_{\mathrm Q}^{2}\over r^{2}})}
\eqno(21)$$
With our assumption regarding the form of $\phi(r,t)$ , this leads to
$$(\dot\phi_{\mathrm 2n}(t))^{2} 
={1\over{\sqrt 3}} + (1- {r_{\mathrm s}\over r}+{r_{\mathrm Q}^{2}\over r^{2}})^{2}(\phi_{\mathrm 1n}'(r))^{2} = k
\eqno(22)$$
where $k$ is a constant. For $k={1\over{\sqrt 3}}$, again  $\phi_{\mathrm 1n}'(r)=0$
and we can write the solution for the $k-$essence field as a homogeneous field
$\phi (r,t)=  d_{1} + (3)^{-1/4}t + d_{2}$, where $d_{1,2}$ are constants.

These configurations again satisfy the emergent gravity
equations of motion $(4b)$ as before:
${L_{X}^{2}\over c_{\mathrm s}}[G^{00}\partial_{0}^{2}\phi_{\mathrm 2n} 
+ G^{11}(\partial_{1}^{2}\phi_{\mathrm 1n} 
-\Gamma_{11}^{1}\partial_{1}\phi_{\mathrm 1n})
+G^{01}\nabla_{0}\nabla_{1}\phi 
+G^{10}\nabla_{1}\nabla_{0}\phi]= 0$. The first two terms within third brackets
vanish because $\phi_{\mathrm 2n}$ is linear in $t$ and $\phi_{\mathrm 1n}$ is
a constant. The last two terms vanish because $G^{01}\nabla_{0}\nabla_{1}\phi 
+G^{10}\nabla_{1}\nabla_{0}\phi=G_{01}\nabla^{0}\nabla^{1}\phi 
+G_{10}\nabla^{1}\nabla^{0}\phi$ and $G_{01}= G_{10}=0$.

The emergent metric in the case of Reisner-Nordstrom background is then :
$$G_{\mu\nu} = \left(\begin{array}{cccc}
{1\over 3} & 0 & 0 & 0\\
0 & {-1\over (1- {r_{\mathrm s}\over r} + {r_{\mathrm Q}^{2}\over r^{2}})^{2}} 
 & 0 & 0\\
0 & 0 & -r^{2}\over (1- {r_{\mathrm s}\over r}+ {r_{\mathrm Q}^{2}\over r^{2}}) & 0\\
0 & 0 & 0 & -r^{2} sin^{2}\theta\over (1- {r_{\mathrm s}\over r}+ {r_{\mathrm Q}^{2}\over r^{2}})\\
\end{array}\right) \eqno(23)$$

\section{ "Masking" }

We now briefly discuss the "masking" of the singularity
(we ignore the singularity at $r=r_{\mathrm s}$ which is a
coordinate singularity as already mentioned in the beginning).
In our treatment, the scalar field is a homogeneous scalar field ,
i.e., depends only on the time coordinate. As the spatial part of the scalar field
is always a constant in $r$ and the temporal part is linear in the temporal coordinate,
our scalar field is well behaved at the central (physical) singularity, {\it viz.}, $r=0$.

The perturbations of the scalar field travels in $G_{\mu\nu}$.
This metric $G_{\mu\nu}$ is perfectly well behaved at $r=0$ as can
be easily seen. Hence in this metric one can never "see" 
the physical singularity. This is what we  mean by  "masking" of the
singularity at $r=0$.

Does this mean that we have done away with the singularity. The answer is
obviously {\it no} as we now show.
For illustrative purposes we shall confine ourselves to the
Schwarzschild case. Let us define $\delta g_{\mu\nu}= G_{\mu\nu} - g_{\mu\nu}$.
Then it is easy to see that
$$\delta g_{00}= {r_{s}\over r}-{2\over 3}~;~ \delta g_{11}={-r_{s}r\over (r-r_{s})^{2}}$$
$$\delta g_{22}={-r_{s}r^{2}\over (r-r_{s})}~;~\delta g_{33}={-r_{s}r^{2}sin^{2}\theta\over (r-r_{s})}\eqno(24)$$
For the Reisner-Nordstrom background the above equations take the form:
$$\delta g_{00}= {r_{s}\over r}-{r_{\mathrm Q}^{2}\over r^{2}}-{2\over 3}~;~ 
\delta g_{11}={-r_{s}r^{3}+r_{\mathrm Q}^{2}r^{2}\over (r^{2}-rr_{s}+r_{\mathrm Q}^{2})^{2}}$$
$$\delta g_{22}={-r_{s}r^{3}+r_{\mathrm Q}^{2}r^{2}\over (r^{2}-rr_{s}+r_{\mathrm Q}^{2})}~;~
\delta g_{33}={(-r_{s}r^{3}+r_{\mathrm Q}^{2}r^{2})sin^{2}\theta\over (r^{2}-r_{s}r+r_{\mathrm Q}^{2})}\eqno(25)$$
Note that in both the above examples the change in the original metric $\delta g_{\mu\nu}$
still carries the same singularity structure at $r=0$ as $g_{\mu\nu}$. This is as it should be.
Therefore the singularity is still there but it is impossible to be aware of it
if we use $G_{\mu\nu}$. This is what we call "masking".

\section{Conclusion}

In this work we have shown that for observers whose world line is in an
emergent gravity metric $G_{\mu\nu}$,

(a)The physical singularity at $r=0$ in the gravitational
metric $g_{\mu\nu}$ can remain masked for certain configurations of the
$k-$essence field $\phi$ and observers travelling with the perturbations of such $k-$essence
fields will never be aware of the physical singularity of
the gravitational metric as this is not conformally equivalent to the emergent gravity metric.

(b)These configurations are homogeneous (i.e. functions of time $t$ only ) and satisfy the
equations of motion in the emergent gravity metric.

(c)The above have been shown here for the Schwarzschild and the Reissner-Nordstrom metrics.

Back reaction effects and inhomogeneous field configurations will be discussed in future communications.

\end{document}